# Engineering Magnetic States and Magnetoresistance in Bisegmented Co-Ni Jellyfish Nanowires via Interplay of Shape and Magnetocrystalline Anisotropies

M.I. Sobirov[1*], K.A. Rogachev[1], M.A. Bazrov[1], Zh. Zh. Namsaraev[1], I.M. Sapovskii[1], T.R. Rakhmatullaev[1], N.V. Ilin[1], A.O Lembikov[1], S.M. Pisarev[1], A.V. Ognev[1,2], A.S. Samardak[1,2], A.Yu. Samardak[1]

[1]Institute of High Technologies and Advanced Materials, Far Eastern Federal University, Vladivostok 690922, Russia.
[2]Sakhalin State University, Yuzhno-Sakhalinsk 693000, Russia

*sobirov.mi@dvfu.ru

**Abstract.** Expanding the spectrum of 3D magnetic nanostructures requires mastering the interplay between different anisotropy contributions. Here, we fabricate bisegmented jellyfish nanowires with tailored arrangements of Co (strong magnetocrystalline anisotropy) and Ni (dominant shape anisotropy) segments. We uncover a unique magnetic duality: Co segments can be tuned to exhibit either a flux-closing multidomain state or a shape-anisotropy-dominated vortex configuration, directly governed by their geometry. This local domain structure, imaged by MFM and simulated micromagnetically, dictates the global magnetic response—suppressing magnetostatic interactions in arrays and enabling the programming of the anisotropic magnetoresistance (AMR) in single nanowires. Our work provides a blueprint for designing functional magnetic nanomaterials where magnetoresistive properties are engineered through strategic anisotropy control.

## 1. Introduction

While the magnetic phenomena induced by two-dimensional shape of ferromagnetic objects and observed in thin films and their derivatives, has found many applications and already serve humanity for decades, the expansion of the nanomagnetism into the third dimension of space can be beneficial due to the many reasons, including novel magnetic phenomena and domain states, more rational use of space and materials, cheaper and faster process of manufacturing [1, 2]. One-dimensional (1D) nanostructures, such as nanowires [3], segmented nanowires [4], nanotubes [5, 6], nanocoils [7] and jellyfish nanowires [8], are possessing strong uniaxial shape anisotropy induced by their high aspect ratio, which can be manipulated by adjusting their spatial curvature, making this class of nanostructures promising candidates for exploring of novel magnetic phenomena, which can not be observed in traditional 2D materials. A number of non-trivial spin configurations was already studied or proposed in such nanostructures – vortices [9], Bloch point domain walls [2], skyrmion tubes [10], hopfions [11] and "corkscrew" states [12] with promising applications.  in transmission and processing devices [4, 13], logical elements of neuromorphic computing[14], drug carriers in targeted drug delivery, agents for local hyperthermia and magnetic cell separation[15], highly sensitive magnetic sensors [16].

The combination of a strong uniaxial shape with other anisotropy types could result in even more exotic spin configurations and open new ways for precise tuning of the magnetic behaviour of magnetic nanostructures. The potent magnetocrystalline anisotropy of hexagonal close-packed Co (hcp) is uniaxial and is orientated according to the c-axis in the unit cell. In the polycrystalline nanomaterials, the orientation of grains and their c-axes can be manipulated comparably easily by change of polycrystalline texture, which can be done by control over conditions of electrodeposition in a various ways, including modification of electrolyte's pH [17-19], voltage of electrodeposition [20, 21], change of the concentration of $Co^{2+}$ ions [21], and application of external magnetic field [22]. Resulting magnetocrystalline anisotropy, in dependance on its direction, can assist shape anisotropy, favoring orientation of magnetization along the long axis of nanowires, or compete with it, creating complex micromagnetic configuration induced by compound local anisotropy [23]. In the latter case, it can change the character of magnetostatic interactions in the array or even overcome the shape anisotropy, which can be useful if the induced shape anisotropy of the domain structure is not appropriate for specific applications [4].

This work is an extension of our previous study, devoted to the preparation and investigation of magnetic properties of Ni jellyfish (JF) nanowires [8] and develops the statements therein in a further direction. During the preparation of the samples, we implemented the suggested method of precise control over the material of the "heads" and "legs" of JF nanowires, resulting in the preparation of bisegmented jellyfish (BSJF) nanowires with Co and Ni in the corresponding parts. We used Co as the material for JF and BSJF nanowires to investigate the influence of the combination of different shape anisotropies and magnetocrystalline anisotropies on the magnetic behaviour of the nanostructures. By combination of experimentally analyzed morphological, structural and magnetic properties, we were able to assume the roles of the magnetocrystalline and shape anisotropies in different segments of obtained nanostructures, while micromagnetic simulations allowed us to confirm our suggestions and relate the observed magnetic phenomena to the domain structure of the samples and dominant type of anisotropy in corresponding segments. As a result, we have managed to carry experimental measurements of the magnetoresistance of the individual nanostructures and implement a methodology of calculation of anisotropic magnetoresistance based on micromagnetic simulation to outline the origins of experimentally obtained data.

## 2. Results and discussion

### 2.1 Sample preparation, morphology, composition and structure

Bilayered nanoporous anodic alumina templates (BNAATs) were prepared by the modified aluminium anodisation method [8, 24]. All of the templates used for the synthesis of jellyfish nanowires (JF) and bisegmented jellyfish nanowires (BSJF) were prepared under the same conditions to ensure that their geometrical parameters are the same and that differences in the magnetic behaviour of the samples are strictly induced by their composition. Detailed conditions for the preparation of BNAATs are described in the Methods section.

Scanning electron microscopy (SEM) images of both surfaces of one of the BNAATs used to synthesize nanowires, are shown in Figure S1 (a-b) in Supplementary FIJI [24] software was used to calculate the geometrical parameters of BNAATs from obtained SEM images. For the top layer, the median diameter was calculated as $D = 173\pm23$, the interpore distance as $D_{int} = $

245±31 nm; for the bottom layer – $D = 87±11$ nm, interpore distance $D_{int} = 120±16$ nm. The normal distributions of the pore diameters in the corresponding layers are shown in Figure S1 (c-d).

Three samples were prepared on the basis of the obtained BNAATs – Co JF nanowires and two bisegmented samples with different orders of Co and Ni – Co/Ni BSJF and Ni/Co BSJF nanowires by implementing two-bath electrodeposition. Detailed conditions of electrodeposition are described in the Methods section. As was proposed in previous work [8], the moment of filling of the pores in the top layer was observed from the current-time dependence graph, to achieve a sharp transition from one material to another in the corresponding layers. SEM images of all synthesized samples are presented in Figure 1 (a-c). The geometric parameters of the obtained samples are presented in Table 1. Following the name from the previous article, we will refer to the part of the JF and BSJF nanowires with a higher diameter from the top layer as "heads" and to another part located in the bottom layer with small diameters as "legs". The differences in the lengths of the "heads" and "legs" in different samples should not significantly affect their magnetic properties, due to the strong shape anisotropy of nanowires with high aspect ratio [26]. The results obtained were compared with the JFS3 sample from previous work [8], as they have comparable geometric parameters. In this work, this sample will be referred to as Ni JF.

Table 1. Geometrical and material parameters of the obtained samples

| Sample | Material | | $D$, nm | | $l$, μm | |
|---|---|---|---|---|---|---|
| | "Heads" | "Legs" | "Heads" | "Legs" | "Heads" | "Legs" |
| Co JF | Co | Co | 292± 11 | 91 ± 7 | 1.99±0,03 | 10.89±0,28 |
| Co/Ni BSJF | Co | Ni | 296 ± 20 | 104 ± 9 | 2.18±0,05 | 10.55±0,42 |
| Ni/Co BSJF | Ni | Co | 298 ± 8 | 95 ± 7 | 1.95±0,04 | 10.96±0,18 |
| Ni JF [8] | Ni | Ni | 273±26 | 99±17 | 1.95±0.04 | 10.32±0.64 |

The study of the elemental composition was carried out by energy dispersive X-ray spectroscopy (EDX) using an EDAX detector integrated into SEM. The results of the EDX analysis for the BSJF nanowires are shown in Figures 1 (d) and (e). Because the control over the deposited material was carried out on the basis of the current-time dependance during electrodeposition, we were able to achieve the uniform filling of the pores in corresponding layers with corresponding materials. Implementation of a two-bath method allowed one to achieve Co and Ni segments in their pure form, without the formation of the CoNi alloy.

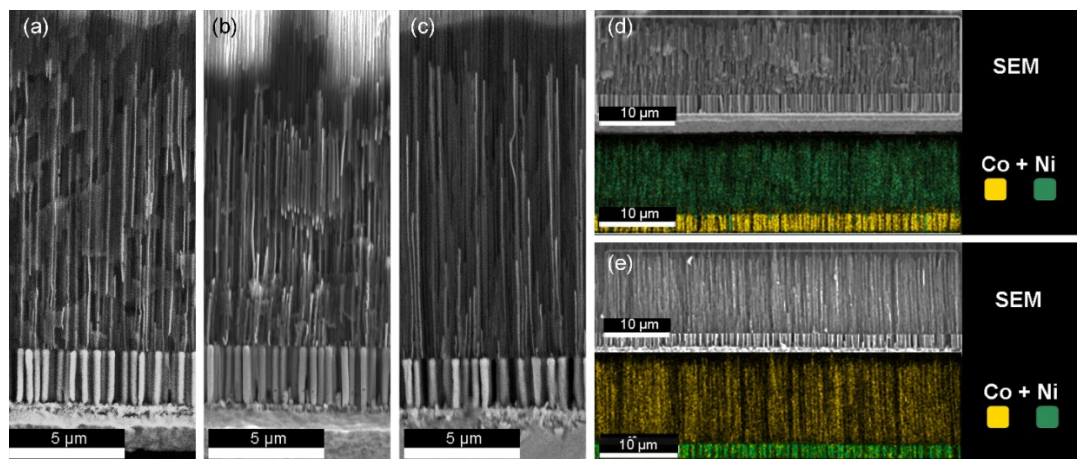

Figure 1. SEM image of arrays of (a) JF Co, (b) BSJF Co/Ni, and (c) BSJF Ni/Co nanowires on the cross section of the BNAAT. SEM and EDX mappings of (d) BSJF Co/Ni and (e) BSJF Ni/Co.

To analyze the crystal structure of the synthesized samples, X-ray diffraction was used. Based on the data obtained (Figure 2), it was found that all samples were polycrystalline, with Co presented in its hexagonal (hcp) form, while Ni was characterized by a face-centered cubic (fcc) lattice. Co in all samples was textured because of the strong difference in the relative reflex intensities compared to the theoretical values, while Ni does not exhibit a strongly preferred orientation. The medium grain size was calculated according to Sherrer's formula and was determined as 40 nm for Co and 25 nm for Ni. Cu signal originated from the conductive layer on the bottom of BNAATs.

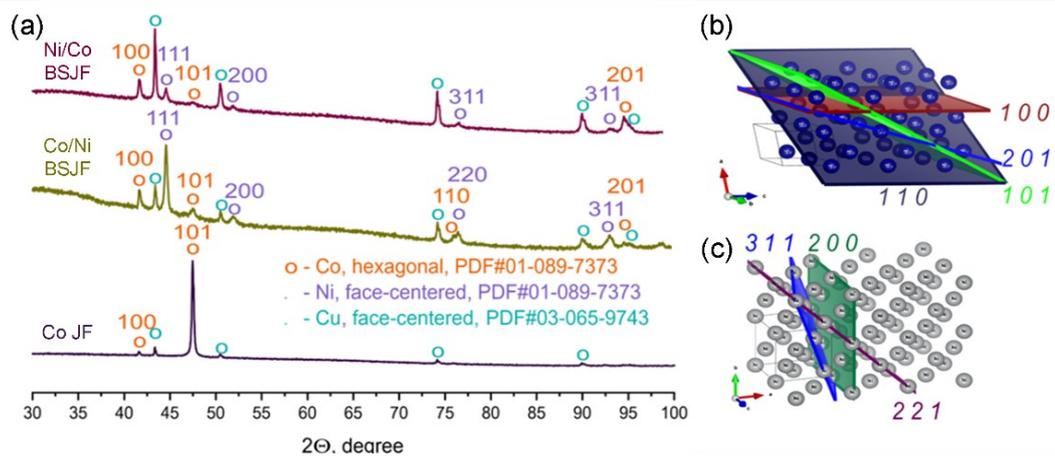

Figure 2. (a) Diffractograms of the corresponding samples. Graphical representation of Co (b) and Ni (c) atomic structures.

## 2.2 Magnetic properties and FORC method

For a detailed analysis of the magnetic properties, nanowire arrays embedded in BNAATs were examined using a vibrating sample magnetometer (VSM), measuring the magnetization $m$ in the external magnetic field $\mathbf{H}$, orientated parallel and perpendicular with respect to the long axis of the nanowires. The captured hysteresis loops are presented in Figure 3, and their magnetic properties are summarized in Table 2.

The resulting hysteresis loops appear clearly distinct from each other, despite having the same geometry of the nanowires across the samples. As explained in detail in our previous work [8], the magnetic properties of Ni JF nanowires (Figure 3 (d)) are defined by their shape anisotropy and magnetostatic interactions, since the magnetocrystalline anisotropy of polycrystalline fcc Ni is insignificant. As simulation and experimental investigation showed, the Ni JF nanowire arrays carry intense interwire magnetostatic interactions between the single domain "legs", which switch in parallel $\mathbf{H}$ by jumping between the parallel and antiparallel directions of $m$. The "heads" of Ni JF nanowires, magnetostatically coupled to their neighbours as well, contain a distorted vortex configuration, named the "corkscrew" state [12]. Intense interactions in the array result in a sloped hysteresis loop in the direction of the easy axis of magnetisation parallel to the long axis of the nanowires (Figure 3 (d)).

Among the other samples with Co, the Co JF nanowires (Figure 3 (a)) appear with the lowest effective anisotropy with an easy direction in $\mathbf{H}$ parallel to the nanowires. Since coaxial direction of shape and magnetocrystalline anisotropies could favour more pronounced effective anisotropy, we could assume that this close to isotropic behaviour can be induced by magnetocrystalline anisotropy, orientated by texturing primarily in the perpendicular direction with respect to the shape anisotropy, which makes these anisotropies compete and decrease resulting effective anisotropy. While such behavior is rather common for textured hcp Co nanowires [19, 20, 23], "heads" and "legs" should have different contributions of shape anisotropy due to their different geometrical parameters, thus resulting effective anisotropy could vary in different parts of the nanowire. From the hysteresis loops of this sample, it is hard to assume the predominant type of anisotropy in the "heads" and "legs", but analysis of the following ones, especially BSJF nanowires, can shed some light on this matter.

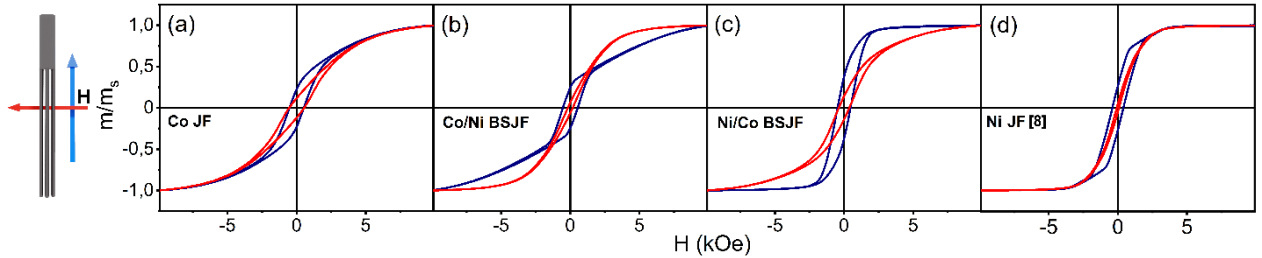

Figure 3. Magnetic hysteresis loops of arrays of (a) Co JF, (b) Co/Ni BSJF, (c) Ni/Co BSJF and (d) Ni JF [8] nanowires, embedded in a BNAAT in the **H** parallel (blue line) and perpendicular (red line) to the long side of the nanowires.

The hysteresis loops for Co/Ni BJFS (Figure 3 (b)) are the most remarkable among the others. In comparison with Co JF nanowires, anisotropy appears to be much stronger in this sample, yet at first glance the direction of the easy axis of magnetisation can be unclear. In **H**, parallel to the long axis of the nanowires, the presence of Co and Ni, as two phases with distinct magnetic properties, manifests itself as a hybrid hysteresis loop, combining both easy and hard directions of magnetisation. One phase appears to saturate in a comparably small external magnetic field $H_{sat}$ = 1.6 kOe, clearly displaying the easy axis of magnetisation in this direction of the **H**. The other phase saturates in fields much higher than $H_{sat}$ = 14 kOe, which greatly exceed the saturation fields of the hysteresis loop in the perpendicular **H** ($H_{sat}$ = 11 kOe). This hysteresis loop can be imaginably dismantled into two simple loops, which would be similar to two extremes of the Stoner–Wohlfarth model [25] in parallel and perpendicular **H**. The first loop, associated with the first phase, is of square shape with a slight slope and rather high coercive force, so we could assume that magnetisation switches there by jumping from the parallel to antiparallel direction, in the range of fields, shifted by a strong magnetostatic interaction. As defined in the previous paper [8], such magnetisation switching occurs in the Ni "legs" due to their high shape anisotropy, so it is safe to assume that Ni "legs" is the first phase with smaller $H_{sat}$. The second loop is characterised by high slope with minimal $H_c$, resembling not switching but coherent rotation of the magnetisation vectors **m** after **H**. It can occur only if the shape anisotropy is not dominant in the system, and the effective anisotropy is orientated perpendicular to the applied **H**. Thus, we can assume that in Co "heads" contribution of the shape anisotropy to the effective anisotropy is minimal, while magnetocrystalline anisotropy is defining an easy axis of magnetisation in the perpendicular direction to the long axis of the nanowires.

The Ni/Co BSJF nanowires (Figure 3 (c)) are characterized by a common shape for 1D nanostructures of the hysteresis loops [26, 27], with pronounced anisotropy and direction of the easy magnetization axis along the long axis of the nanowires. This shape is signalling about the effective anisotropy orientated along the long axis of the nanowires and the dominance of the shape anisotropy in both Ni "heads" and Co "legs". Thus, we could assume that in all samples investigated, the magnetic properties of Co "heads" are conditioned mostly by magnetocrystalline anisotropy, while the magnetic behavior of Co "legs" is defined by shape anisotropy.

As presented in Table 2, the Ni JF sample is characterized by the lowest $H_c$ and highest ratio of remnant magnetization to saturation magnetization $m_r/m_s$. The highest $H_c$ is observed in Co/Ni BSJF nanowires, which is slightly higher than in Co JF, both of which have a comparable $m_r/m_s$ ratio. This observation suggests that the presence of Co does increase $H_c$, but its specific combination with Ni increases it even more.

Table 2. Magnetic characteristics of all samples, extracted from hysteresis loops and FORC diagrams.

| Sample | $H_c$, (∥) Oe | $H_c$, (⊥) Oe | $H_{sat}$, (∥) kOe | $H_{sat}$, (⊥) kOe | $m_r/m_s$ (∥) | $m_r/m_s$ (⊥) | $H_c^{FORC}$, (∥) Oe | $H_u$, (∥) Oe |
|---|---|---|---|---|---|---|---|---|
| Co JF | 515 | 547 | 13 | 14 | 0.19 | 0.12 | 0-969 | 734 |
| Co/Ni BSJF | 535 | 173 | 14 | 11 | 0.22 | 0.06 | 260-881 | 722 |
| Ni/Co BSJF | 462 | 378 | 8.5 | 13.5 | 0.36 | 0.13 | 0-981 | 634 |
| Ni JF [8] | 397 | 90 | 3.5 | 4 | 0.5 | 0.06 | 272-540 | 1090 |

For a more detailed study of the magnetic properties of the array, samples were investigated using the FORC method [28, 29], a tool for analyzse systems of magnetically interacting nanostructures, which makes it possible to determine the distributions of interaction fields ($H_u$) and local coercive forces ($H_c^{FORC}$). The FORC diagrams obtained with the external magnetic field **H** orientated parallel (∥) to the long axis of the nanowires are presented in Figure 4 (a-d). For a more detailed understanding of the magnetic behavior of the obtained samples, a FORC diagram for the homogenous Ni sample from our previous work [8] was also added to the comparison.

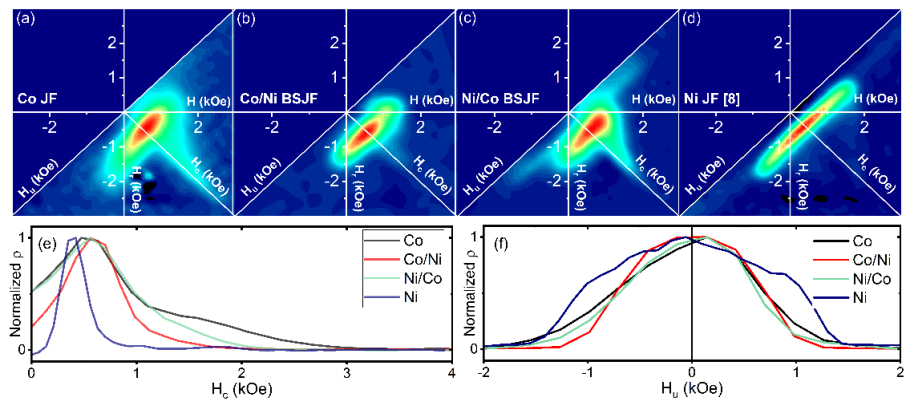

Figure 4. FORC diagrams for (a) Co JF, (b) Co/Ni BSJF, (c) Ni/Co BSJF, and (d) Ni JF [8] nanowires in the direction of **H** parallel to the long axis of the nanowires. Profiles of the FORC distribution along the (e) $H_c$ and (f) $H_u$ axes.

The FORC-diagrams for the Co and Ni/Co samples (Figure 4, (a) and (c)) are characterized by a similar stingray-like shape, with a long ridge of local coercive force $H_c^{FORC}$ distribution along the Hc axis (Figure 4 (e)), while the Co/Ni and Ni samples appear more uniform in terms of $H_c^{FORC}$ (Figure 4, (b) and (d)). The observed $H_c^{FORC}$ values (table 2) for Co and Ni/Co samples, in comparison with the others, suppose that the presence of Co "legs" in the system can both increase local $H_c$, (the upper limit of the $H_c^{FORC}$ distribution reaches 1 kOe), and decrease the local value of $H_c$, (the lower limit of the distribution begins at $H_c^{FORC}$=0 Oe), resulting in a wide range of the overall distribution across the $H_c$ axis. In contrast, the lower value of the $H_c^{FORC}$ distribution for samples with Ni "legs" does not start from 0 values and is instead fixed at the same level $H_c^{FORC}$ = 265 Oe for both samples. The reason for such behavior of $H_c^{FORC}$ can be the influence of magnetocrystalline anisotropy of Co, whose direction in the polycrystal can be distributed chaotically and create points with low and high local coercive force along the nanostructure. For Ni, the local coercive force $H_c^{FORC}$ is sensitive to the shape of the nanostructures, because of lack of other contributors to the effective anisotropy, thus the narrow distribution of $H_c^{FORC}$ reflects uniform geometrical parameters of the sample.

In terms of interaction fields $H_u$, all samples with Co appear with a similar distribution, significantly lower than for a uniform Ni sample ($H_u$ = 634-734 Oe versus $H_u$ = 1090 Oe, respectively). This suggests that introducing Co into a magnetically coupled system can noticeably reduce the intensity of magnetostatic interactions, despite Co having a higher saturation magnetization $M_{sat}$ than Ni (1400 kA/m versus 490 kA/m, respectively [30, 31]). The reason for such, at first glance, deviant behavior could originate from two reasons: the complex domain structure in the Co "legs" and the possible presence of reversible switching in the "heads".

In Co "legs", a complex domain structure induced by a combination of strong shape and magnetocrystalline anisotropies can enclose the magnetic flux within the nanostructure and decrease magnetostatic interactions. Or it can direct it out of the structure and strongly increase the interaction. The latter case was analyzed in our previous work for polycrystalline hcp Co nanowires with a smaller diameter $D$ = 40 nm and a grain size of 40 nm [23], which showed a significant increase of magnetostatic interactions in the array, due to the orientation of the magnetization in each grain perpendicular to the direction of the nanowire long axis . In this work, the diameter of the "legs" is greater that a grain size ($D$~100 nm versus a grain size of 40 nm), which means that some number of grains can compose the "leg" across its diameter. Random directions of the c-axes of these grains orientated by texture in plane, perpendicular to the long axis of the nanowires, can favour the closure of magnetic flux inside the nanowire, and accordingly decrease magnetostatic interactions in the array. Nevertheless, from the measurements of the hysteresis loops we observed that shape anisotropy should dominate in the Co "legs" which should create significant interactions between closely packed nanostructures in the array. Thus, we could expect some middle-ground situation, where orientation of **m** in Co "legs" is dominated by shape anisotropy, yet with some micromagnetic configuration that favours the enclosure of magnetic flux inside their volume, without the creation of single-domains like in their Ni counterparts.

In contrast, the weak shape anisotropy of the "heads" can result in the direction of effective anisotropy perpendicular to the long axis of the nanowire, making the parallel direction into a hard magnetization axis. In this case, according to Stoner–Wohlfarth model, reversible mechanisms of remagnetization, such as coherent rotation, should prevail, making them undistinguishable for FORC method, since it takes information from Barkhausen jumps induced by irreversible domain wall motion [29]. Therefore, even if magnetostatic interaction is presented between the "heads", it cannot be observed using FORC, and we should investigate this matter by means of other methods. However, the FORC-diagram of the Ni/Co BSJF sample (Figure 4 (d)) in addition to the main distribution contains a very weak, but very broad distribution along $H_u$, which was observed in the previous work and was attributed to interactions of Ni "heads" [8].

Magnetic force microscopy (MFM) of singular JF nanowires, extracted from BNAAT, revealed unexpected and drastically different characters of the stray field distributions of Ni and Co JF nanowires (Figure 5). The first one is depicted with 3 points of strong magnetic flux – one massive white pole on the end of the "head", one black pole on the opposite end of the "head" and another black pole – at the end of the "legs". However, the Co JF nanowire, despite having higher

$M_{sat}$, is characterised by a very weak magnetic response, without formation of poles with an intensity comparable to those observed in the Ni JF nanowire. Such data confirm the suggestion on closure of magnetic flux closed primarily inside the volume of Co segments in JF and BSJF nanowires. Such big differences between Ni and Co made the MFM analysis of the BSJF nanowires non-informative, since the Ni signal was the only visible signal, and thus their MFM images were excluded from the comparison. Nevertheless, closely packed weak magnetic poles dislocated on both the "head" and "legs" in the Co JF nanowire are visible, which can indicate the presence of a complex domain structure, induced by combination of shape and magnetocrystalline anisotropies.

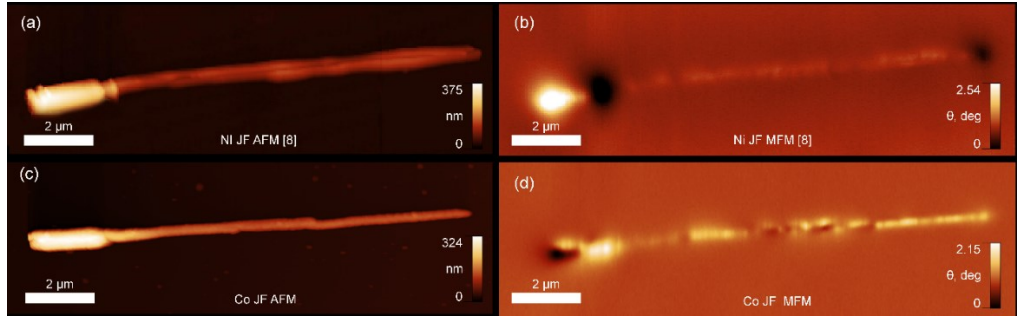

Figure 5. AFM and MFM of (a, b) Ni JF [8] and (c, d) Co JF nanowire, respectively.

## 2.3 Micromagnetic simulations

To obtain a detailed picture of the magnetic behavior of arrays and singular JF and BSJF nanowires at the micro level and build a base for understanding their magnetoresistance, micromagnetic modeling was performed using MuMax$^3$ [32]. The model parameters are described in detail in the "Methods" section.

To achieve reliable simulation results, comparison of the simulated and experimental hysteresis loops for each sample was made as shown in Figure 6. The simulation of arrays was carried out for 12 nanowires, ordered in hexagonal order. The simulated hysteresis loops demonstrate good qualitative and quantitative agreement with the experimental data for all samples at both orientations of the **H**, with minor deviations due to the inevitable simplifications of the micromagnetic model. Additionally, simulated hysteresis loops for singular nanostructures were also added to the comparison in Figure 6. Interestingly, the single Co JF and Co/Ni BSJF nanowires (Figure 6, (a) and (b) red dotted line) were characterised by the same shape of the hysteresis loops, compared to the results of the array, but their slope was much steeper with significantly lower $H_{sat}$. As proposed during FORC analysis, this signifies strong magnetostatic interactions between the "heads" in the array, which cannot be detected by FORC due to the reversible origin of their magnetization switching.

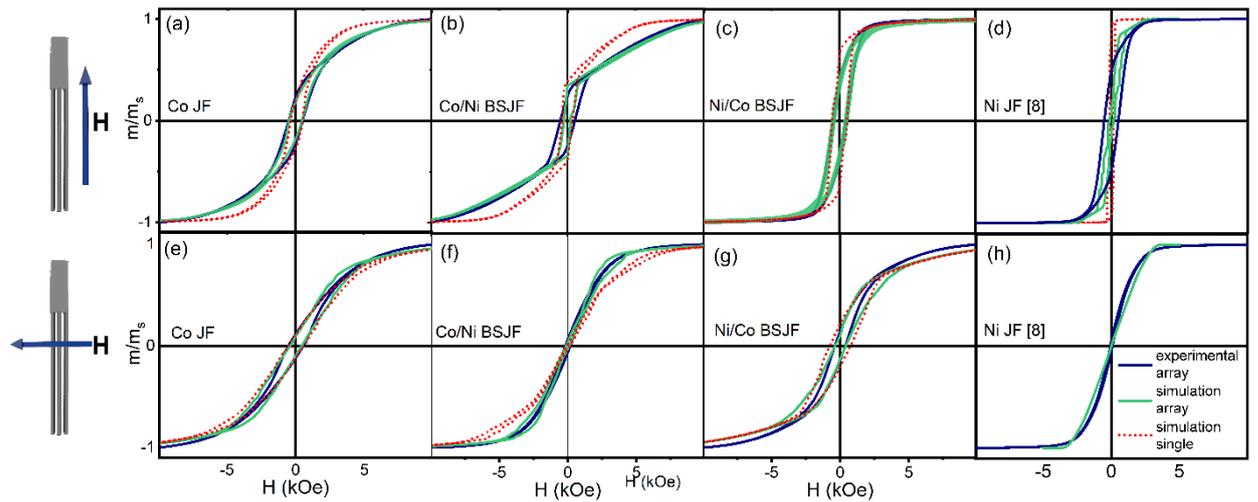

Figure 6. Comparison of experimental (blue line obtained from the array) and simulated (green line from the array and red dotted line from the single nanostructure) magnetic hysteresis loops in **H** applied (a, b, c and d) parallel and (e, f, g and h) perpendicular to the long axis of nanowires.

The results of the visualization of magnetization **m** and its angular velocity for all samples in the remnant state after saturation in parallel **H** are presented in Figure 7. Dynamics of nanostructures switching is presented in the Supplementary movie SM1 for all samples and discussed in detail in the next section since its close connection with magnetoresistance. The domain structure of the Co and Ni segment was quite similar in different samples, reporting a minor influence of neighbouring segments on their behaviour. The Co "heads" of Co JF and Co/Ni BSJF nanowires were characterized by a complex multidomain structure, distorted by the influence of magnetocrystalline anisotropy of the individual grains, with **m** orientated primarily perpendicular to the long axis of the nanowires while magnetic flux tended to mostly close inside their volume. In contrast, the Ni "heads" in Ni JF and Ni/Co BSJF nanowires appear to have a single-domain configuration, with vortices as the end domains. The Co "legs" in Co JF and Ni/Co BSJF samples have a set of subsequent vortices of different chirality, with domain walls between them distorted by means of magnetocrystalline anisotropy of Co grains. As predicted from experimental magnetic measurements, such presence of vortices is a conscience of the domination of the shape over magnetocrystalline anisotropy, since most of the **m** is aligned along the long axis of nanowires. Noticeably, while the "corkscrew" configuration was presented in the "head" of the Ni JF nanowire during remagnetization, it was not observed in the Ni/Co sample, which can describe this micromagnetic configuration as the product of specific interactions between the "head" and "legs". In this sample the domain configuration of the "legs" appears similar to the observed Co JF nanowire, yet the period of the domains seems shorter, with a higher number of vortices and domain walls, which can be the consequence of the magnetostatic influence of the Ni "head".

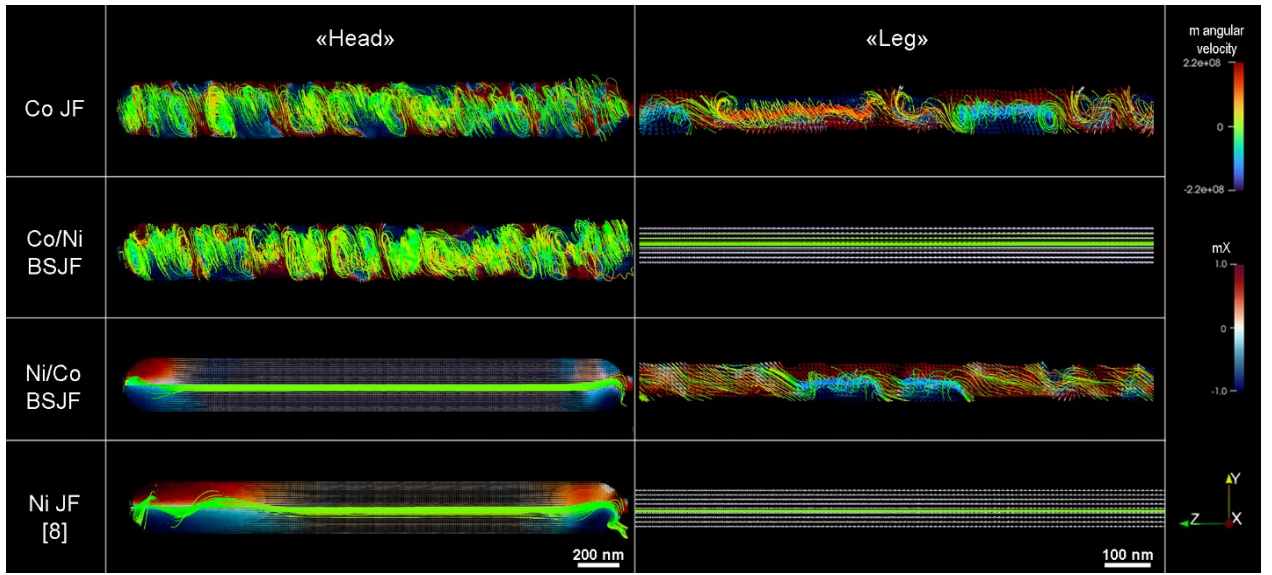

Figure 7. Configuration of magnetization (red-white-blue arrows) and its angular velocity (rainbow lines) in singular "head" and "leg" of the corresponding samples.

## 2.4 Magnetoresistance measurements and simulation

For most of their applications in electronics, metallic nanostructures must be subjected to the current flow. Conductivity of ferromagnetic nanomaterials, in addition to other surface and quantum-induced effects, is influenced by configuration of their spins, inducing effects such as anisotropic magnetoresistance (AMR) [33], giant magnetoresistance (GMR) [34] and tunneling magnetoresistance (TMR) [35]. AMR, while considered very weak in comparison with the other types of magnetoresistance, is still proposed for use in industry for variety of sensors and magnetic memory [16]. For a more detailed study of the behavior of JF and BSJF and the possibility of their use as spintronics devices, we investigated the magnetotransport properties of single nanowires with respect to their simulated magnetic configuration.

The setup for AMR measurements for all samples is presented in Figure 8. In addition to the Co JF, Co/Ni and Ni/Co BSJF samples, the magnetotransport properties of homogeneous Ni JF nanowires from our previous work [8] were studied. Direct current was applied through the contacts 1 and 4, and voltage was measured at sensor contacts 2 and 3 (Figure 8 (e)), with the nanowire placed in the region with a dashed circle. The micromagnetic configuration as a vector matrix was obtained by simulation in MuMax3 and was used to calculate the dependence of the resistance of the samples in an external magnetic field using a self-written Python program that calculates the conditional resistance by measuring the angle of deviation of the magnetization of each discretization cell from the direction of the current flow [36]. A detailed description of the parameters of AMR measurements and calculations is given in the Methods section.

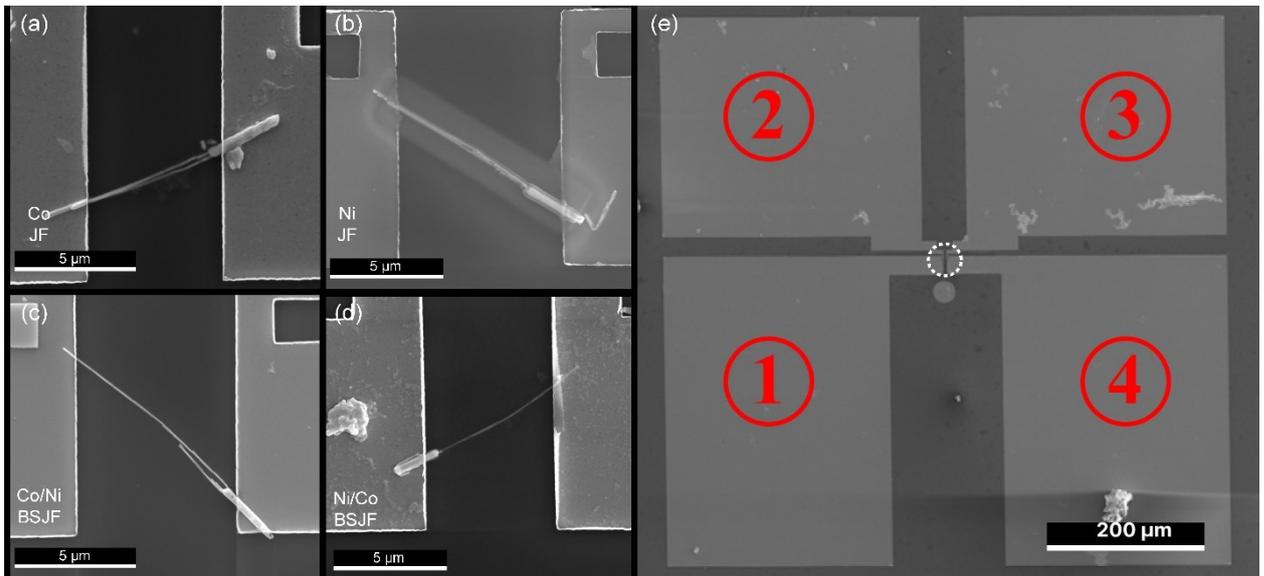

Figure 8. SEM images of (a) Co JF, (b) Ni JF, (c) Co/Ni BSJF, and (d) Ni/Co BSJF placed between the contact pads for AMR measurements. (e) General view of contact pads.

The dependencies of resistance $\Delta R/R$ on **H** for all samples are presented in Figure 9. The designations "Up" and "Down" on the resistance charts indicate changes **H** from minimum to maximum and vice versa, respectively. The coercive force is understood as the maximum value of $\Delta R/R$ in the perpendicular **H** and the minimum drop at the peak in the parallel **H**. While the experimental and simulated curves were qualitatively similar, quantitatively there were inevitable differences due to the presence of contributions to resistance, which cannot be considered in micromagnetic simulations such as defects, shape deviations and Joule heating. The parallel **H** was orientated in a direction from the "head" to the "legs", and the transverse **H** was in the plane of the substrate. To increase the trustworthiness of the results, the geometry of the nanowires in the simulation was adjusted to match their real counterparts.

In general, all samples show a similar dependence of $\Delta R/R$ for the same orientation of **H**: The increase in **H** parallel to the long axis of the nanowire direction increases resistance, and the increase in **H** in the perpendicular direction decreases resistance. This AMR behavior is classical for 1D ferromagnetic nanostructures and is associated with the angle between the current flow and spin orientations. In all samples the value of $\Delta R/R$ did not exceed 1%, which is typical for both hcp-Co [37] and Ni [38]. The highest AMR value of 1% was observed for perpendicular direction of **H** for Co/Ni BSJF sample (Figure 9 (f)).

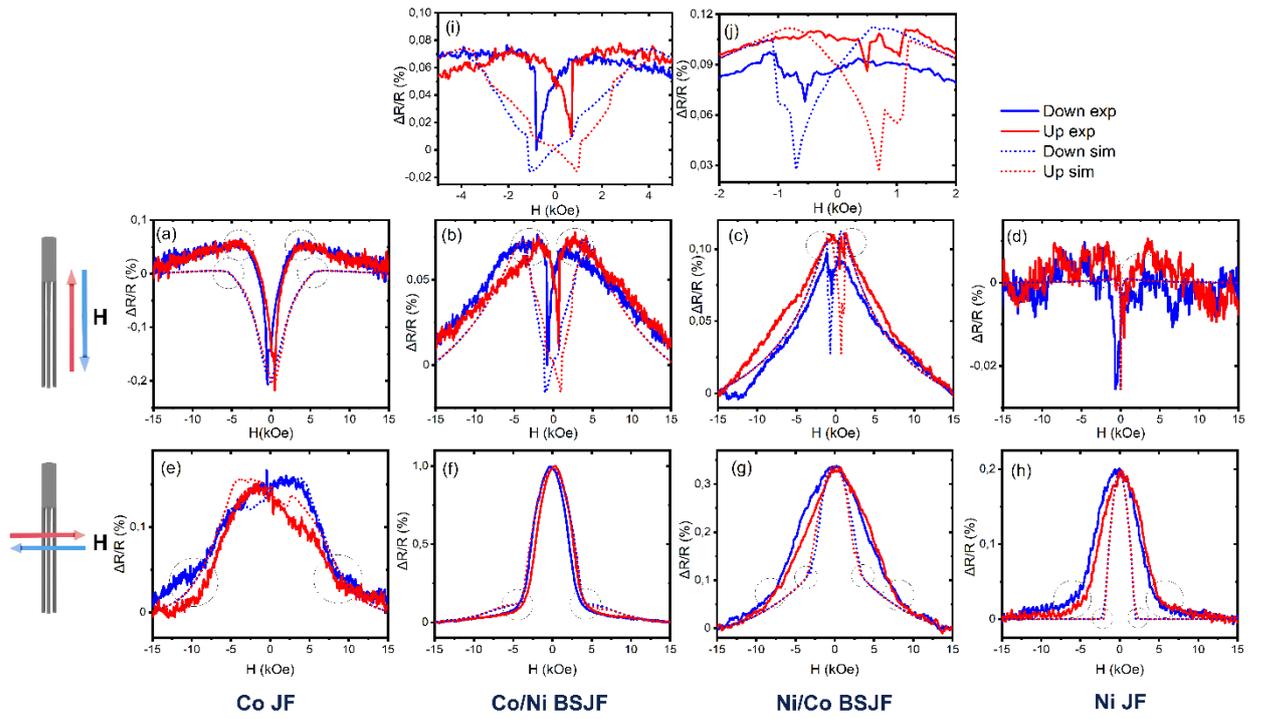

Figure 9. Dependence of resistance Δ*R*/*R* of a single nanowire on **H**, orientated in parallel (a-d, i, j) and perpendicular (e-h) direction for samples (a,e) Co JF, (b, f, i) Co/Ni BSJF, (c, g, j) Ni/Co BSJF, and (d,h) Ni JF. Graphs (i) and (j) are enlarged versions of (b) and (c), respectively.

The slope of the Δ*R*/*R* curves in **H** parallel to the main axis of the nanowire, observed on the experimental graphs for all samples except Ni JF (Figure 9 (a-c)), occurs due to the presence of an angle between the **H** and long axis of the nanowire, the non-parallelism of the axis of the nanowire's "head" and the axis of its "legs", as well as the curved shape of the "legs". As in the simulation shown (Figures S2-S5 in the Supplementary) such an angle between **H** and the long axis of the nanowire deflects the direction of magnetization from the direction of current flow, reducing resistance. In this case, the maximum resistance (marked with circles on the graph) in the parallel **H** is achieved when the magnetization of the entire nanowire is orientated along the direction of current flow (∥$H_a$ and ∥$H_{-a}$ for all of the samples in Figures S2-S5). The complex micromagnetic configuration of the samples with different shapes and magnetocrystalline anisotropy, as well as different interactions of "heads" and "legs" was deciphered by implementation of the calculation of AMR based on the micromagnetic configuration of the samples.

As simulation showed (Figures S2-S5), Co "heads" and "legs" are switching in **H**, parallel to the long axis of the nanowire, by the formation and subsequent annihilation of multiple magnetic domains along their entire length, which is induced by the presence of strong magnetocrystalline anisotropy. For the Ni "heads" and "legs", the switching mechanisms differ significantly from each other and their Co counterparts: in Ni "heads" switching occurs by volumetric formation of vortices, while "legs" switch by formation and movement of the domain walls, maintaining a single domain configuration prior and after switching events due to high shape anisotropy.

The small differences in the switching of "heads" and "legs" of the same compositions in different samples in **H**, parallel to the long axis of the nanowire, can be explained by the presence of an adjacent element made of a different material. In the Co/Ni BSJF nanowire (Figure S3), the process of switching the "head" still occurs by the formation and subsequent annihilation of multiple magnetic domains throughout its entire volume, but the number of these domains is significantly reduced, and they acquire a more pronounced orientation along the main axis of the nanowire, which can be conditioned by magnetostatic interaction with single domain Ni legs. In the Ni/Co BSJF nanowire (Figure S4), the influence of the neighboring segment is more pronounced, volumetric vortices do not form in the Ni "head", and it switches by the formation of vortices at its ends, which may be due to the weak magnetostatic interaction with the Co "legs" compared to the Ni JF nanowire. However, in the Co "legs" of this sample, switching occurs by the formation of domains along the entire length, however, in the presence of a Ni "head", a magnetic domain with a predominant orientation along the long axis is also formed and propagates by the influence of increasing **H**, which can originate the magnetostatic interaction between the "head" and the "legs". When the field changes in the other direction, such a domain does not form, which may be due to the asymmetry of the wire geometry. Unexpectedly, no "corkscrew" configuration was observed for samples with the Ni/Co "head", which could be a consequence of different material composition of the "legs", inducing different magnetostatic interactions and breaking this fragile micromagnetic configuration.

Since in parallel **H** the Ni "legs" in Ni JF and Co/Ni JFS switch between single domain states with the **m** direction parallel or antiparallel to the current flow, their influence on AMR should be neglected due to its tolerance to such changes of **m** orientations. In contrast, the "head" of Ni JF nanowire should greatly influence resistance due to the creation of a "corkscrew" state, decreasing the AMR, since a significant volume of the "head" is occupied by **m**, orientated perpendicular to the long axis of the nanowire and the vector of the current flow. Thus, the resistance drop in Figure 9 (d) could be associated mainly with the vortex domain structure of the Ni "head", with some neglectable additions from the domain walls in the "legs".

In the Ni/Co BSJF nanowire the drop in resistance is bifurcated due to two mechanisms of resistance decrease in the system, that occurred in different **H**. First, one could be associated with creation of rather big vortex end domains. The other one – with the creation and annihilation of multiple domain walls with a perpendicular direction of magnetization with respect to the current flow in multidomain configuration of the Co "legs". In the Co/Ni BSJF only Co "heads" should primarily influence the resistance, since Ni "legs" appear in the single domain state; thus the observed drop in the AMR curve is due to the turn of **m** in a plane, perpendicular to the long axis of the nanowires. In the Co nanowires, the most significant resistance drop is observed, since both the "head" and "legs" have a large portion of **m**, orientated perpendicular to the long axis of the nanowires.

In perpendicular **H**, coherent rotation of **m** in the direction perpendicular to the long axis of the nanowires occurs in most samples, which results in a steep decrease of resistance with an increase of **H**. The only exception is Co "heads", which possesses magnetization switching in low perpendicular **H**, which does not influence magnetoresistance due to indistinguishability of the resulting directions of **m** for current.

### 3. Conclusions

In conclusion, we have successfully engineered and characterized a new class of functional magnetic nanomaterials: bisegmented Co-Ni jellyfish nanowires. By strategically combining segments with dominant shape anisotropy (Ni) and strong magnetocrystalline anisotropy (Co), we demonstrate precise control over the effective magnetic anisotropy on a segment-by-segment basis. This anisotropy engineering directly dictates the resulting complex domain structures, which we unravel through a combination of advanced characterization techniques and micromagnetic simulations.

Key to this control is the texturized hcp-Co structure, whose random in-plane c-axis orientation competes with shape anisotropy. This competition leads to unique magnetic responses, including the unusual coexistence of easy and hard axis behaviors within a single hysteresis loop for the Co/Ni configuration. Crucially, we establish that the complex, flux-closing domain structures in Co segments—visually confirmed by MFM—significantly reduce magnetostatic interactions within arrays, a critical factor for device integration. Furthermore, we directly link these engineered micromagnetic states to functional properties by measuring and simulating the anisotropic magnetoresistance of individual nanowires, providing a blueprint for designing magnetoresistive response at the nanoscale.

This work fundamentally advances the field of 3D nanomagnetism by providing a clear pathway to program magnetic behavior through the sophisticated design of material composition and anisotropy. The ability to tailor domain configurations and magnetostatic interactions in these nanowires makes them highly promising candidates for next-generation applications, particularly in high-density spintronic memory and neuromorphic computing systems, where tunable magnetic states and low interwire interference are paramount.

**Methods**

*Preparation of bilayered nanoporous anodic alumina templates*

BNAATs were synthesised by two-stage aluminium anodising [8, 24] using various acids. All templates were prepared under the same conditions to ensure similarity. Before anodising, the plates, made of pure aluminium (99.995%, Girmet LLC) with diameters of 25 mm and a thickness of 0.5 mm, were electrochemically polished in 1:4 solution of $HClO_4$ and $C_2H_5OH$ using an Agilent 6030A power supply for 60 seconds at a current density of 500 mA/cm$^2$, Cu plate was used as the cathode. Anodizing of the first layer of BNAATs was carried out in potentiostatic mode using a 5% $H_3PO_4$ solution cooled to 0-1 °C at a potential of 150 V in a home-made anodization cell. To reduce the barrier layer at the bottom of the first layer the anodizing, voltage was gradually reduced from 150 to 120 V by consistent increase of temperature of the solution, while power supply was switched to galvanostatic mode. After that, the samples were immersed in a solution of $Cr_2O_4$ + 5% $H_3PO_4$ heated to 45 °C for 20 minutes to reduce the barrier layer. The anodizing of the second layer was carried out for 6 hours using a solution of 0.3 M $C_2H_2O_4$ in a potentiostatic mode at 60 V. After the anodization process, the non-oxidized aluminum was etched from the back of the plate using a solution of 0.08 M CuCl and 8% HCl for 40 minutes. To obtain BNAATs with through pores, the plates were immersed in $Cr_2O_4$ and 5% $H_3PO_4$ solution heated to 45 °C for 15 minutes.

*Preparation of jellyfish and bisegmented jellyfish nanowires*

To create a conductive layer for electrodeposition, the surface of the first layer of the BNAATs was coated with a 500 nm thick Cu layer sputtered using a homemade thermal evaporation system. To ensure the best conductivity, the deposited Cu layer was thickened by electrochemical deposition using a solution of CuSO4 (20 g/100 ml) + H2SO4 (6 g/100 ml) under a potential of 0.3V for 10 minutes on the opposite to the BNAAT side of the Cu film.

As noted in [8], we have observed that during electrochemical deposition of material into BNAATs value of current or voltage is different during filling of the first and second layers due to different contact area between electrolyte and conductive material, what made it possible to control the filling of BNAATs with different materials in corresponding layers. After a rapid change of current was detected during the deposition of the first layer, the process was stopped, and the sample was moved to an electrochemical cell with another solution to obtain BSJF nanowires with different segment compositions.

Each of the samples was placed in an electrochemical cell as a cathode with the corresponding electrolyte at room temperature: $CoCl_2×6H_2O$ + $H_3BO_3$ was used for Co electrodeposition with Co plate as anode; $NiSO_4×6H_2O$ + $NiCl_2×6H_2O$ + $H_3BO_3$ was used for Ni electrodeposition, with Ni plate as anode. JF Co nanowires were electrochemically deposited for 70 minutes using a 0.6 V potentiostatic mode by Keithley SourceMeter 2460. For Co/Ni BSJF nanowires, Co was electrodeposited for 26 minutes at a potential of 0.6 V, filling the first layer of BNAAT, with consequent deposition of Ni for 200 minutes at a potential of 0.9 V into the second layer of BNAAT. Ni/Co BSJF nanowires were electrochemically deposited for 75 minutes at a potential of 0.9 V to fill the first layer with Ni and 30 minutes at a potential of 0.6 V to deposit Co into the second layer.

*Morphological, elemental and structural analysis*

The morphology of porous alumina templates and nanowire arrays on cross-section of BNAATs, as well as nanowires etched out of the template, was studied using the ThermoScientific SCIOS 2 two-beam system with 10 kV voltage and current in the range of 50 pA - 0.4 nA. The built-in EDX (EDAX, TEAM) detector was used to perform elemental analysis.

The crystal structure of nanowires embedded in BNAAT was studied at room temperature using a Kolibri diffractometer (IC Burevestnik). The Cu anode of the X-ray tube (Cu-Ka radiation) was connected to a linear Muthen2 detector in the Bragg-Brentano geometry. The crystal structure was refined using the Rietveld method using TOPAS 4.2 software. The weighted deviation coefficient ($R2$) in the calculations did not exceed 7%. The experimental diffractograms were approximated using the Fityk software. The crystal size was calculated using the Scherrer formula.

*Magnetic measurements*

Hysteresis loops and FORC diagrams for nanowire arrays embedded in a BNAAT were measured using a LakeShore VSM 7410 vibration sample magnetometer at room temperature.

The FORC diagrams were measured with the direction of the external magnetic field **H** along and perpendicular to the long axis of the nanowires at a maximum of $H = \pm 8$ kOe with a step change of $H_r = 100$ Oe, resulting in a set of 160 inverse curves for each diagram.

The surface relief and stray field distribution were studied using an atomic force microscope (AFM, NT-MDT NTEGRA Aura) with a magnetic CoCr needle MFM01 (NT-MDT NTEGRA) with a tip curvature radius of 40 nm. Scanning of Co JF nanowires was performed at a distance of 20 nm from the surface by a contactless method with preliminary removal of the surface relief using the same probe.

*Micromagnetic simulations*

Micromagnetic simulations were performed with MuMax3 [32], which solves the Landau–Lifshitz–Gilbert (LLG) equation employing the finite difference method for space definition. Calculations were performed with a Nvidia Tesla P100. The 5 nm cell size was chosen according to the magnetic exchange length of Co, which allowed the resolute of nanowires and Co grains in the model, as well as the simulation of the nanostructure array in a volume of $1100 \times 1480 \times 12700$ nm$^3$. Periodic boundary conditions were established at the boundaries of the simulated region in a plane perpendicular to the long axis of the nanowires to neutralize the boundary effects. Following the results of the morphological analysis, the diameter of "heads" and "legs" was set at 300 and 100 nm for the Co JF and Ni/Co BSJF nanowires, respectively. The length of "heads" and "legs" was set at 2 μm and 10.5 μm, respectively. With the exception of the length of the "legs" in all samples and the length of the "heads" in the Co/Ni nanowires, no other geometric parameter of the nanowires and their array were changed from sample to sample in accordance with the use of aluminium oxide templates with the same morphology.

The saturation magnetization $M_{sat}$ was set to 1.3 and 0.35 MA/m for the Co and Ni segments, respectively, which is lower than the values for bulk materials [31] due to the polycrystalline structure of the nanowires, as well as the increased presence of impurities and defects common for electrochemical deposition. The exchange stiffness $A_{ex}$ was set to 30 and 9 PJ/m for Co and Ni, respectively. All samples were divided into grains with size 40 nm following the results of structural analysis using 3D Voronoi tessellation. In each Co grain, a random direction of the axes of magnetocrystalline anisotropy was set in the plane perpendicular to the long axis of nanowires. To recreate the polycrystalline texture of hcp Co, the length of the anisotropy vector was limited to 20–40% along the long axis of the nanowire, while in the perpendicular directions its length was set to 100%. For Co, the uniaxial magnetocrystalline anisotropy was set as $K_{u1}$ = 520 kJ/m3.

The simulation results were visualised using ParaView.

*Magnetoresistance measurements and simulation*

To study the magnetotransport properties of individual nanostructures, 120 nm thick W contact pads were sputtered on the surface of oxidized Si templates with oxide thickness of 400 nm using electron beam lithography (ThermoScientific SCIOS2) and magnetron sputtering (Omicron). To prevent oxidation and improve heat dissipation from the measured sample, a layer of non-conductive silicon oxide was applied to the nanowires using focused electron beam induced deposition (FEBID, ThermoScientific SCIOS2). A four-probe method was implemented, a constant current was supplied through current contacts 1 and 4, and the voltage was measured at sensor contacts 2 and 3 (Figure 8), and the change in resistance in **H** from (+15) kOe to (-15 kOe) was studied on a self-made magnetoresistance measuring system. A direct current of 150 to 200 μA was supplied by a laboratory Keithley 6221 power source.

The micromagnetic configuration as a vector matrix was obtained by simulation in MuMax3 [32]. It was used to calculate the dependence of the resistance of the samples in an external magnetic field using a self-written Python program that calculates the conditional resistance by measuring the angle of deviation of the magnetization of each cell from the direction of the current flow [36].

**Acknowledgements.** The study was carried out with the support of the Russian Science Foundation (project No. 24-72-10088). Ognev A.V. thanks the Ministry of Education and Science of Russia for support under a state assignment (project No. FZNS-2023-0012). The work involved equipment of an integrated common use centre and interdisciplinary research in the field of nanotechnologies and new functional materials (Far Eastern Federal University, Vladivostok, Russia).

**Supplementary files**

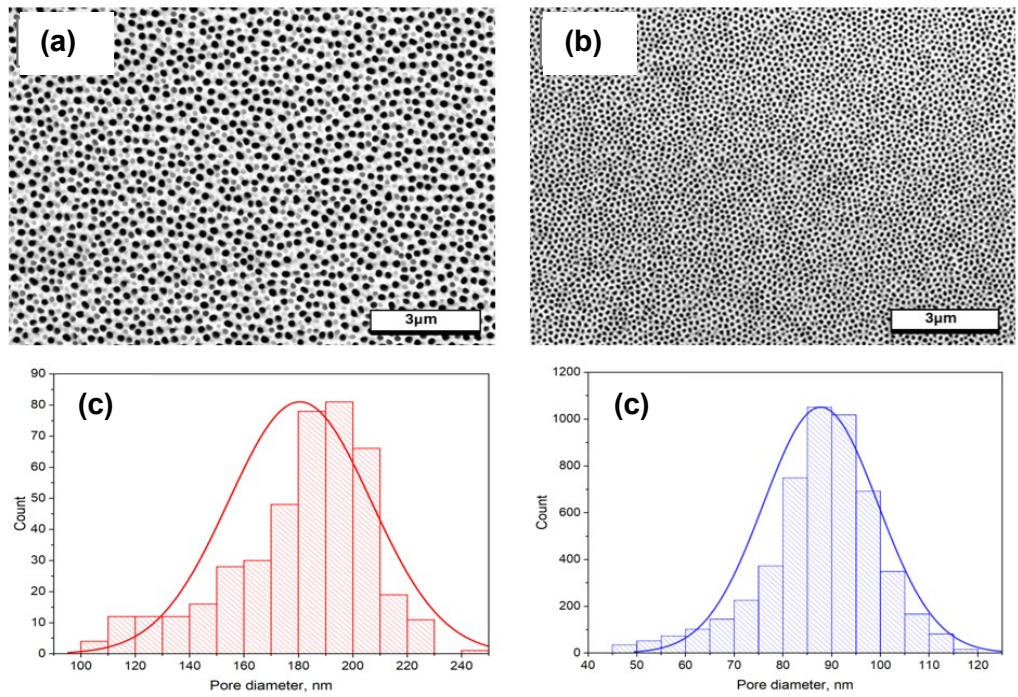

Figure S1. SEM images of BNAAT used for the synthesis of the samples, (a) top and (b) bottom surfaces, (c) and (d) distribution of diameters in corresponding layers.

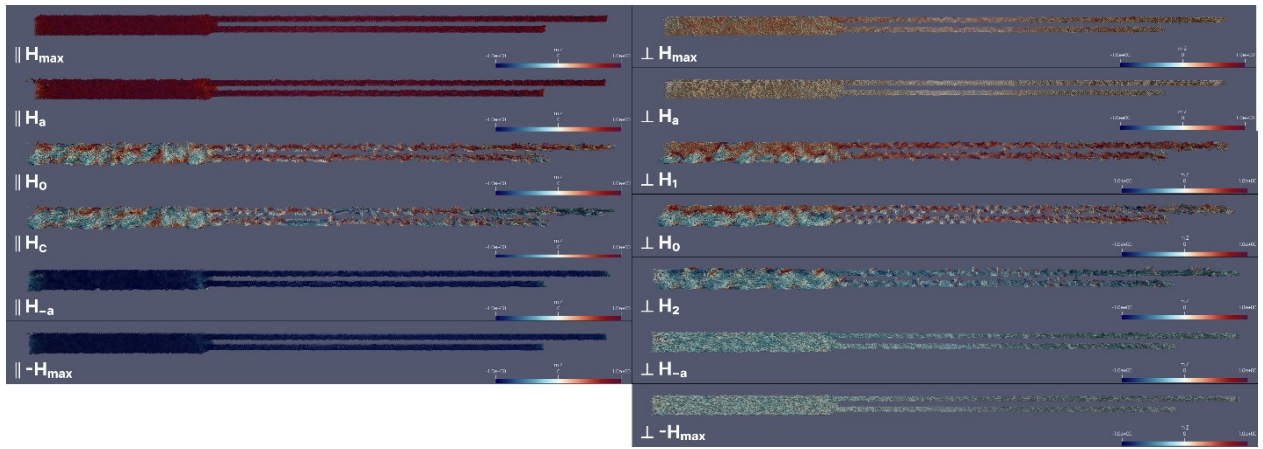

Figure S2. Visualization of the magnetic configuration of the Co JF nanowire during remagnetization in the transverse (left column) and longitudinal (right column) **H**

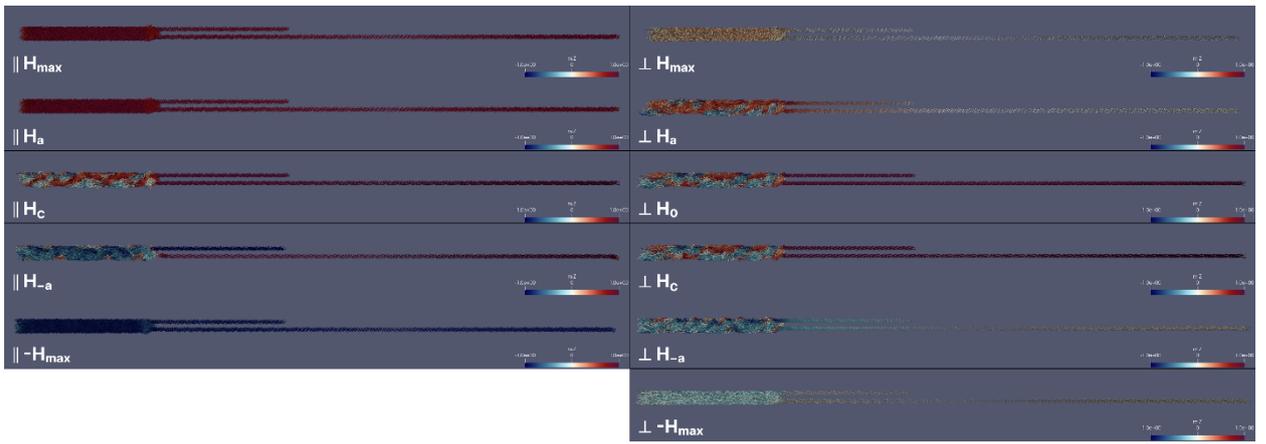

Figure S3. Visualization of the magnetic configuration of the Co/Ni BSJF nanowire during remagnetization in the transverse (left column) and longitudinal (right column) **H**

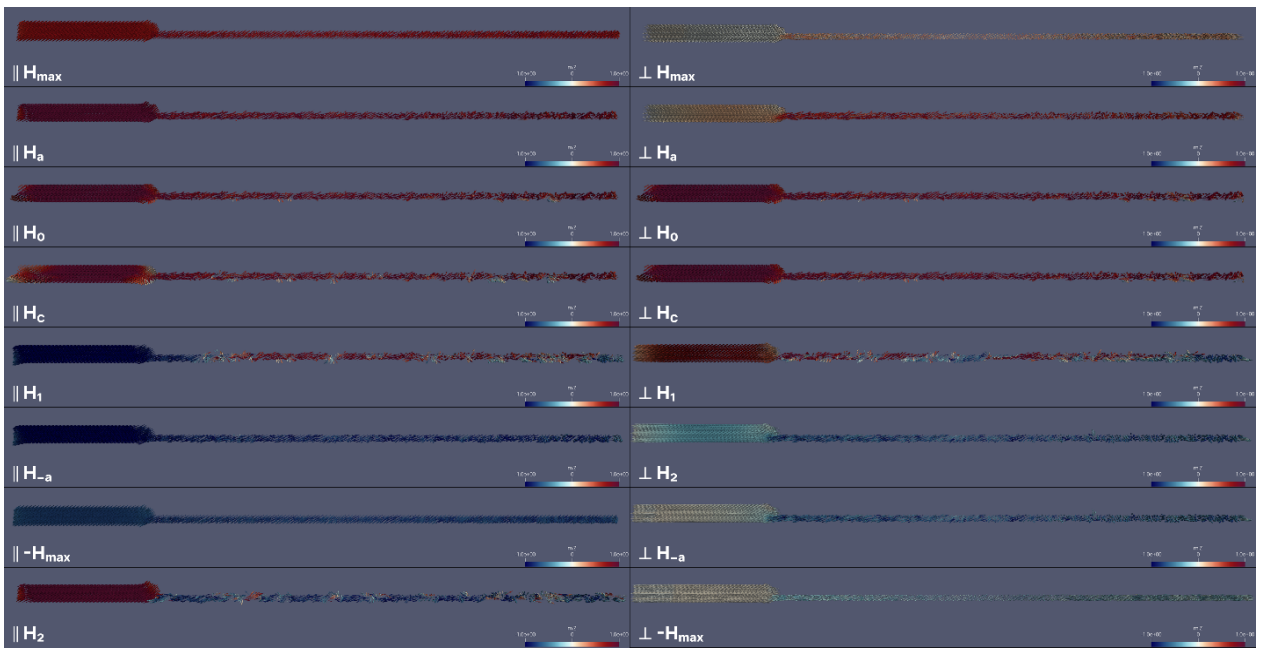

Figure S4. Visualization of the magnetic configuration of the Ni/Co BSJF nanowire during remagnetization in the transverse (left column) and longitudinal (right column) **H**

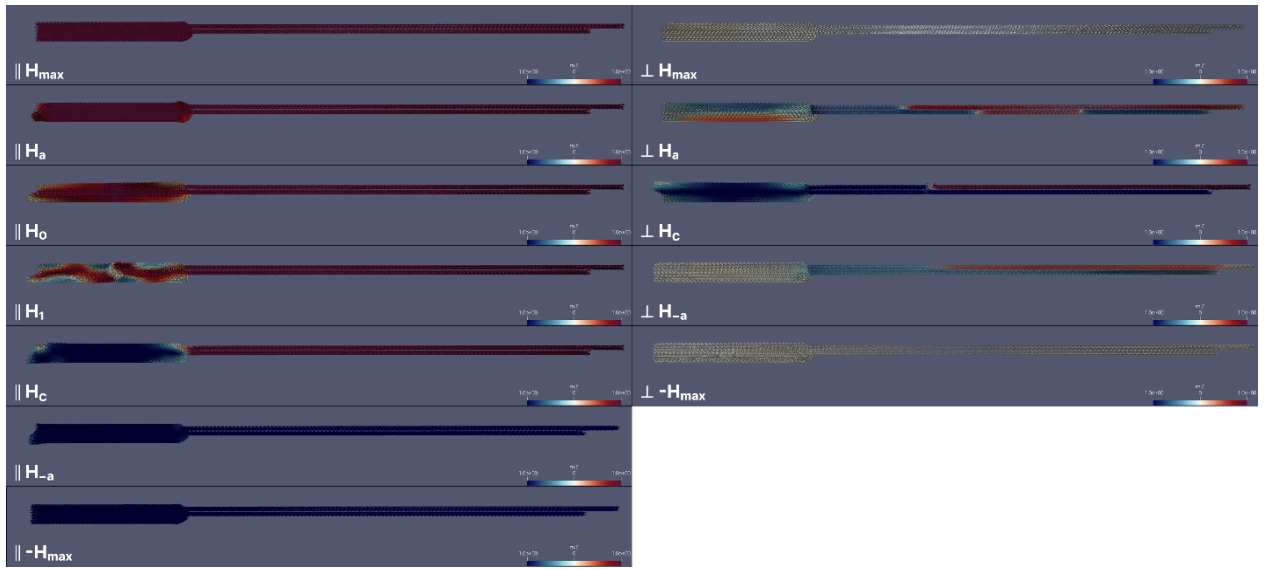

Figure S5. Visualization of the magnetic configuration of the Ni JF nanowire during remagnetization in the transverse (left column) and longitudinal (right column) **H**